\documentclass[iop,apj]{emulateapj}
\pdfoutput=1 
\usepackage{apjfonts}
\usepackage{mathtools}
\usepackage{color}
\usepackage[colorlinks, citecolor=blue, pdfborder={0 0 0}]{hyperref}

\newcommand{\msun}{M_{\odot}}
\newcommand{\sn}[2]{#1{\times}10^{#2}}

\shorttitle{Deep Pulse Search in LMXBs}
\shortauthors{Patruno et al.}

\begin{document}

\title{A Deep Pulse Search in Eleven Low Mass X-Ray Binaries}

\author{A. Patruno}
\affiliation{Leiden Observatory, Leiden University, P.O. Box 9513, NL-2300 RA Leiden, The Netherlands}
\affiliation{ASTRON, Netherlands Institute for Radio Astronomy, Postbus 2, NL-7990 AA Dwingeloo, The Netherlands}

\author{K. Wette}
\affiliation{Max Planck Institute for Gravitational Physics (Albert Einstein Institute) and Leibniz Universit\"at Hannover, D-30167 Hannover, Germany}
\affiliation{ARC Centre of Excellence for Gravitational Wave Discovery (OzGrav) and Centre for Gravitational Physics, Department of Quantum Science, Research School of Physics and Engineering, The Australian National University, Canberra ACT 2601, Australia}

\author{C. Messenger}
\affiliation{SUPA, School of Physics and Astronomy, University of Glasgow, Glasgow G12 8QQ, United Kingdom}

\begin{abstract}
  We present a systematic coherent X-ray pulsation search in eleven low mass
  X-ray binaries (LMXBs). We select a relatively broad variety of LMXBs,
  including persistent and transient sources and spanning orbital periods
  between 0.3 and 17 hours. We use about 3.6 Ms of data collected by the
  \textit{Rossi X-Ray Timing Explorer} (\textit{RXTE}) and \textit{XMM-Newton} and apply a semi-coherent search strategy to look for weak and persistent pulses in a wide spin frequency range. We find no evidence for X-ray pulsations in these systems and consequently set upper limits on the pulsed sinusoidal semi-amplitude between 0.14\% and 0.78\% for ten outbursting/persistent LMXBs and 2.9\% for a quiescent system. These results suggest that weak pulsations might not form in (most) non-pulsating LMXBs.
\end{abstract}



\section{Introduction}

An important question in the study of neutron star low mass X-ray binaries
(LMXBs) is why most of them do not show accretion powered pulsations. Only a
small fraction of them has measurable pulsations, with typical pulsed amplitudes
of the order of 1-10\%, that reveal the spin frequency of the neutron star. The
spin is a key quantity to measure because it is related to a number of important
fundamental physics and stellar astrophysics problems, like the equation of
state of ultra-dense matter~\citep{lat07}, the evolution of the neutron star
magnetic field~\citep{bha91} and allows tests of general relativity in strong
gravity via pulse profile modelling~\citep{mor11}. Those systems that show
pulsations with spin periods in the millisecond range are classified in two
broad categories:
\begin{itemize}
\item accreting millisecond X-ray pulsars (AMXPs), with accretion powered
  pulsations (19 systems known to date, see ~\citealt{pat12r,str17});
\item nuclear powered X-ray pulsars (NXPs), with burst oscillations seen during
  thermonuclear bursts (10 systems known to date, beside a few AMXPs which are
  also NXPs;~\citealt{gal08b, wat12}).
\end{itemize}
The reason why some bursting LMXBs show burst oscillations is not completely
understood and it is currently believed that this might be related to the
physical conditions at the ignition point on the neutron star surface~(see
e.g.,~\citealt{gal17}). If a neutron star LMXB has a magnetic field of the order
of $10^8$ G or more, it should display X-ray pulsations since the field is
sufficiently strong to channel the gas towards the magnetic poles. The fact that
most neutron star LMXBs are not AMXPs is therefore not understood. There are
several models that attempt to explain the lack of pulsations, but they all come
with weaknesses that seem incompatible with the growing body of observational
results collected so far (see \citealt{pat12r} for a detailed discussion).

An important aspect of this conundrum is that the current non-detection of
pulsations might be simply ascribed to the presence of very weak pulses which
are below the sensitivity of current instrumentation and/or current search
techniques. All AMXPs have indeed been discovered so far by simply looking at
the power spectra and by identifying the spin frequency by eye inspection (see
for example~\citealt{wij98, mar03, alt10, alt11}).  The fractional amplitude of
their pulsations reaches values of 20--30\%~\citep{pat10,alt11}, whereas the
smallest pulsed fractions observed so far are around
$\sim1$\%~\citep{gal07,pat09e}.

A complication in this scenario is the existence of three so-called intermittent
AMXPs~\citep{gal07, gav07, cas08, alt08} which show pulsations only sporadically
during their outbursts. The mechanism behind their intermittent behavior is
still not known. Furthermore, more sophisticated attempts to detect weak pulses
in the neutron star LMXB 4U 1820--371~\citep{dib04} and Aql X-1 ~\citep{mes15}
have led to negative results, with upper limits on the pulsed fraction of less
than $\approx 0.3$--$0.5$\%.

Apart from these rare exceptions, upper limits smaller than $\approx 1$\% on the
pulsed fraction are not available beside a few of the brightest LMXBs (the
so-called Z sources, see \citealt{vau94}).  Therefore it is important to push
the current upper limits to smaller values, since there is no a-priori reason to
believe that LMXBs should not be able to form pulsations with fractional
amplitudes smaller than $1$\%. In this work we thus investigate this problem more
systematically than done in the past.

We select eleven accreting neutron stars in LMXBs with different orbital
parameters spanning a relatively large range, in order to avoid the selection of
a specific sub-type of LMXBs or a specific evolutionary stage of the binary. We
then apply a semi-coherent search strategy, first developed by \citet{mes11} and
then applied to the source Aql X-1 by \citet{mes15}. To do so we use archival
\textit{Rossi X-ray Timing Explorer} (\textit{RXTE}) data collected over the
lifetime of the instrument, as well as \textit{XMM-Newton} data.  We present the
selection of LMXBs in Section~\ref{sec:selection-lmxbs}, the data reduction in
Section~\ref{sec:obs}, the details of the semi-coherent search strategy in
Section~\ref{sec:sem} and the results of the search in Section~\ref{sec:res}. We
discuss the physical implications of our results in Section~\ref{sec:dis}


\section{Selection of LMXBs}\label{sec:selection-lmxbs}

When searching for pulsations with a semi-coherent search code, it is highly
desirable that the following two criteria are met, in order to allow a deep
pulse search when operating with limited computational resources.  First, the
neutron star should have relatively precise constraints on at least one of its
orbital/spin parameters, in order to reduce the volume of parameter space that
must be searched.  Second, the data need to be relatively closely spaced in
time. The computational cost of a semi-coherent search scales rapidly with the
total \emph{timespan} of the data, whereas the search sensitivity scales
much more slowly with the total \emph{amount} of data contained within the
timespan. Closely-spaced data will therefore maximize the search sensitivity
given a fixed computational budget; for further discussion see~\citet{mes11}
and ~\citet{mes15}.

In order to meet the aforementioned criteria we have looked for LMXBs with
either a robust detection of the orbital period (usually from optical
observations) or sources with a relatively well known spin frequency (thanks to
burst oscillations).  Indeed the purpose of this work is not only to find the
spin frequency of new sources, but also to verify whether weak and persistent
pulsations exist. Therefore sources like 4U 1608--52 and 4U 1636--53, both with
a known spin frequency and with a relatively well constrained orbit, are optimal
candidates for our search.  We have also included the source XTE J1739--2859
despite the lack of any constraint on the orbital parameters, to verify whether
we can find the candidate spin frequency of 1122 Hz reported by~\citet{kaa07}.

To avoid selecting a biased sample of LMXBs with a specific evolutionary and/or
accretion state we have used a mixture of sources, both persistent and
transient, at high and low inclinations and with different orbital periods
corresponding to ultra-compacts (orbital period $P \approx 0.3$~hr) up to
relatively wide binaries ($P \approx 17$~hr). To be conservative in our search,
we have also used a broader parameter space than the formal uncertainties
provided in the literature on the spin/orbit of each source. 
The selection of the parameter space to explore is based on a number of factors, 
like the available computational resources and the robustness of the orbital 
parameters measured in previous works. The semi-major axis
is calculated assuming the most extreme combination of donor and neutron
star masses ($M_{\rm NS}=1.2$--$2.3\msun$). We assume that all binaries have
zero eccentricity and that, aside from one source, the orbital phase is unknown. A summary of the
parameter space explored is given in Table~\ref{tab:par} and a more detailed
description of each source selected is provided in the following subsections.

\begin{table*}
\caption{Spin--Orbit Parameter Space}
\centering
\begin{tabular}{lcccc}
\hline
\hline
Source                & Spin Frequency & Orbital Period & Projected Semi-major Axis & Time of Ascension            \\
                      & $\nu$ (Hz)     & $P$ (s)        & $a$ (lt-s)                & $t_{\mathrm{asc}}$              \\
\hline
4U 1323--619 (wide)   & 50--1500       & 10590--10592   & 0.1745--1.1689            & assumed unknown              \\
4U 1323--619 (narrow) & 50--1500       & 10590--10591   & 0.545--0.633              & assumed unknown              \\
4U 1456--32 (Cen X-4) & 50--1500       & 54000--54720   & 0.04--1.9                 & assumed unknown              \\
4U 1543--624          & 50--1500       & 1073--1111     & 0.00143--0.0599           & assumed unknown              \\
4U 1608--52           & 615--625       & 44064--47521   & 0.3--4                    & assumed unknown              \\
4U 1636--53           & 580--583       & 13655--13656   & 0.35--1.2                 & MJD 50869.00225--50869.02625 \\
XTE J1710--28         & 50--1500       & 11811--11812   & 0.136--0.9                & assumed unknown              \\
4U 1735--44           & 50--1500       & 16746--16748   & 0.05--2.3                 & assumed unknown              \\
XTE J1739--2859       & 1120--1124     & 3600--43200    & 0.01--2                   & assumed unknown              \\
4U 1746--37           & 50--1500       & 18586--18590   & 0.2--1.5                  & assumed unknown              \\
XTE J2123--058        & 50--1500       & 21384--21492   & 0.1--2.45                 & assumed unknown              \\
4U 2129+12 (AC 211)   & 50--1500       & 61603--61608   & 0.39--1.69                & assumed unknown              \\
\hline
\end{tabular}
\label{tab:par}
\end{table*}

\subsection{4U 1323--619}

This is a dipping LMXB with a very well determined orbital period from X-ray
observations~\citep{par89,lev11} and it shows very regular bursts. For this source we
used a 2$\sigma$ interval around the best determined orbital
period of \citet{lev11} but also a much wider parameter space from \citet{par89}.

\subsection{4U 1456--32 (Cen X-4)}\label{sec:4u-1456-32}

4U 1456--32, also known as Cen X-4, is a relatively wide binary with a period of
15.1 hr. It is the only quiescent LMXB in our sample and we used 80 ks of
\textit{XMM-Newton} data collected during March 1, 2003. This is the same
dataset used in \citet{dan15} and we refer to that paper for details. In this
work we have pushed the search to a deeper sensitivity than was done in
\citet{dan15}, who found a 6.4\% upper limit on the pulsed fraction.

\subsection{4U 1543--624}

The persistent LMXB 4U 1543--624 is an ultra-compact binary with an orbital
period of $18.20\pm0.09$ minutes~\citep{wan15}. We used an uncertainty on the
orbital period about 7 times larger than the nominal one.

\subsection{4U 1608--52}

4U 1608--52 is a transient LMXB showing burst oscillations at 619
Hz~\citep{har03} and it is the fastest known spinning accreting neutron
star. The binary orbit is approximately 12.89 hours and it has been determined
from optical variability~\citep{wac02}. However, some ambiguities still exist on
the possibility that the observed variability is due to a super-hump. There is a
very large amount of data recorded by \textit{RXTE} on this source, so we
selected only two outbursts.

\subsection{4U 1636--53}

This is a persistent LMXB with thermonuclear bursts showing burst oscillations
at a frequency of about 581 Hz. Optical data provide a relatively well
constrained orbital period of about 3.8 hours~\citep{ped81} which has been 
refined by VLT observations taken in 2003 \citep{cas06}.

\subsection{XTE J1710--28}

The eclipsing LMXB XTE J1710--28 has a very well constrained orbital period of
about 3.3 hours~\citep{jai11} with a nominal error of about 30 $\mu$s. However,
"glitches" in the mid-eclipse time were detected at the level of a few
milliseconds. We thus used a range about 10 times larger than the glitch size.

\subsection{4U 1735--44}

This is a persistent LMXB and a burster (but no burst oscillations have been
seen), with an orbital period of about 4.6 hours determined from optical
observations of the irradiated donor star and an inclination of 36--60
deg.~\citep{cas06}. The ephemeris are determined with great precision, with a
1$\sigma$ statistical error on the orbital period of only 0.3 seconds. We chose
a wider range of about 3 seconds for our search.

\subsection{XTE J1739--2859}

This is a transient source with unknown orbital parameters. The reason why we
include it in our search is because~\citet{kaa07} reported the detection of
burst oscillations at a frequency of 1122 Hz. This detection has remained, so
far, unconfirmed. However, we are not aware of sophisticated attempts to search
for accretion powered pulsations from this source. Since our limited
computational resources require that the parameter space to search is not too
large, we restricted the candidate orbital periods to values between 1 and 12
hours.

\subsection{4U 1746--37}

The persistent source 4U 1746--37 is located in the globular cluster NGC
6441. It shows bursts (but no burst oscillations) and dips that give an accurate
orbital period of $5.16$~hr~\citep{bal04,lev11}.

\subsection{XTE J2123--058}

This is a transient and a bursting pulsar with no known burst oscillations but a
well determined orbital period of about 5.9 hours from optical spectroscopic
data collected with the Very Large Telescope~\citep{cas02}. The nominal
1$\sigma$ error reported was about 0.2 s and it was obtained by combining the
results with the photometric studies of \citet{zur00}. To avoid any possible
uncertainty due to systematics we used a much broader range of about $\pm50$~s
around the best determined orbital period.

\subsection{4U 2129+12 (AC 211)}

4U 2129+12, also known as AC 211, is located in the globular cluster M15 and its
orbital period of approximately 17 hours, very well determined from X-ray
observations of eclipses~\citep{wen06, ioa02}. The 17 hours orbit implies that
the donor cannot be a main sequence star (that would underfill its Roche lobe)
since the turn off mass of M15 is about 0.8$M_{\odot}$. Furthermore the system
is a peculiar one since it is an accretion disk corona (ADC) source, i.e., the
central source should be permanently obscured by a cloud of material. However,
we included the source in our sample because there is a known ADC source with a
slow accreting pulsar with pulsed fractions of about 1-2\%~\citep{jon01}.


\section{X-Ray Observations}\label{sec:obs}

We used pointed observations collected with the Proportional Counter Array (PCA)
aboard \textit{RXTE} of ten of the eleven LMXBs; for the remaining source Cen
X-4 we used \textit{XMM-Newton} data (see Section \ref{sec:4u-1456-32}). Since
the volume of data recorded is sometimes very large and since we are using
limited computational resources, we selected (for certain sources) only a subset
of the total data available.  A total of $\approx $3.6 Ms of data have been used
in this work.

The data were recorded either as Event ($2^{-13}$ s sampling time) or GoodXenon
($2^{-20}$ s). The GoodXenon data were rebinned by a factor 8192 so to match the
Event time resolution. This speeds up the calculations while still retaining the
necessary narrow pulse sensitivity.  We then retained only photons falling
within the absolute channel range 5--37 ($\sim2$--$16$~keV), which, at least in
known AMXPs, usually maximizes the signal-to-noise ratio of the pulsations. To
avoid that this specific selection of the energy band might bias our search, we
selected a broader energy band, corresponding to absolute channels 5--67
($\sim2$--$30$~keV), for three (arbitrarily chosen) sources.

We then removed all thermonuclear bursts, by defining a burst start and end as
the points in the lightcurves where the X-ray flux becomes twice the pre-burst
level. The data are then barycentered according to the best available ephemeris
(J2000) found in the literature, by using the DE405 JPL Solar System ephemeris.
The source list along with all the program IDs used, the total duration of the
observations, total number of photons collected, absolute channels and the right
ascensions and declinations used are reported in Table~\ref{tab:summary}.

\begin{table*}
\caption{Summary of X-Ray Observations}
\centering
\begin{tabular}{llllll}
\hline
\hline
Source                & Abs. Channels & Right Ascension & Declination   & Program IDs                              & Duration (ks) \\
\hline
4U 1323--619          & 5--37         & 13:26:36.31     & -62:08:9.9    & 20066, 40040, 70050,                     & 339.5         \\
                      &               &                 &               & 90062, 95442, 96405                      &               \\
4U 1456--32 (Cen X-4) & 0.3-10 keV & 14:58:21.92 & -31:40:07.4 & 0144900101 (XMM) & 68.5\\
4U 1543--624          & 5--37         & 15:47:54.29 & -62:34:11.2   & 20064, 20071                             & 39.5          \\
4U 1608--52           & 5--37         & 16:12:43.0      & -52:25:23     & 70058, 70059, 70069                      & 442.0         \\
                      &               &                 &               & 91405                                    &               \\
4U 1636--53           & 5--37         & 16:40:55.57     & -53:45:05.2   & 30053                                    & 49.9          \\
XTE J1710--28         & 5--37         & 17:10:12.3      & -28:07:54     & 40135,40407,60049, 80045                 & 598.3         \\
                      &               &                 &               & 91018, 91045, 93052, 94314               &               \\
                      &               &                 &               & 96329                                    &               \\
4U 1735--44           & 5--67         & 17:38:58.3      & -44:27:00.0   & 10068, 10072, 20084, 30056,              & 1177.9        \\
                      &               &                 &               & 40030,40031,40033, 50025,                &               \\
                      &               &                 &               & 50026, 50029, 60042, 70036,              &               \\
                      &               &                 &               & 91025, 91152, 93200, 93406, 96325, 96331 &               \\
                      &               &                 &               & 96325, 96331                             &               \\
XTE J1739--2859       & 5--37         & 17:39:53.95     & -28:29:46.8   & 91015                                    & 106.9         \\
4U 1746--37           & 5--67         & 17:50:12.7      & -37:03:8.0    & 10112, 30701, 60044-02-*,                & 441.9         \\
                      &               &                 &               & 70050, 90044, 91037                      &               \\
XTE J2123--058        & 5--67         & 21:23:14.54     & -5:47:53.2    & 30511                                    & 66.5          \\
4U 2129+12 (AC 211)   & 5--37         & 21:29:58.3124   & +12:10:02.670 & 10077, 20076, 40041, 92440,              & 374.6         \\
                      &               &                 &               & 95443, 96408, 96428                      &               \\
\hline
\end{tabular}
\label{tab:summary}
\end{table*}


\section{Semi-Coherent Search}\label{sec:sem}

A detailed description of the semi-coherent search strategy used in this work
can be found in~\citet{mes11} with an application to the LMXB Aql X-1
in~\citet{mes15}. Here we briefly summarize the most relevant aspects of the
semi-coherent search useful to understand our results.

\subsection{Method}\label{sec:search-procedure}

The semi-coherent search method comprises two stages. First, in the coherent
stage, the data are partitioned into $M$ short segments of duration $T$; in this
work $T$ ranges from 20 to 3600 seconds. The signal phase\footnote{Note that
Eq.~\eqref{eq:phase} fixes two sign errors with respect to Eq.~(13) in~\citet{mes15}}
\begin{equation}
\label{eq:phase}
\phi(t) = 2\pi \nu \Big[ t - t_0 - a \sin \big( \Omega ( t - t_0 ) + \gamma \big) \Big]
\end{equation}
in the $m$th segment is approximated by a Taylor expansion:
\begin{equation}
\label{eq:deriv}
\phi(t) \approx 2\pi \sum_{s=0}^{s^{*}} \frac{ \nu^{(m)}_{s} }{ s! } ( t - t_0)^s \,,
\end{equation}
where $\nu^{(m)}_{s} \equiv d(\phi(t)/2\pi) / dt |_{t = t_0}$, and $t_0$ is a
reference time.  The search parameters $\nu$, $a$, $\Omega \equiv 2\pi/P$ and
$\gamma \equiv \Omega ( t_0 - t_{\mathrm{asc}} )$ identify the spin frequency,
projected neutron star semi-major axis, orbital frequency and orbital phase,
where $t_{\mathrm{asc}}$ is the time of ascension. Matched filtering of the data
against the model of Eq.~\eqref{eq:deriv} is performed over a search grid in
$(\nu^{(m)}_{0}, \nu^{(m)}_{1}, \dots, \nu^{(m)}_{s^{*}})$.  The highest derivative
order $s^{*}$ ranges from 2 to 4, depending on the value of
$\Omega T$; larger $\Omega T$ require higher-order expansions.

Second, in the incoherent stage, we combine the results of the coherent searches
from each segment.  For each orbital template $(\nu, a, \Omega, \gamma)$, the
derivatives $(\nu^{(m)}_{0}, \nu^{(m)}_{1}, \dots, \nu^{(m)}_{s^{*}})$ are
computed for $m = 1$ to $M$, and the $M$ matched filters corresponding to those
derivatives in the $M$ segments are selected.  Finally the powers in the $M$ matched filters
are summed to give our detection statistic. The number of orbital templates used
in the search scales as
\begin{multline}
\label{eq:nrandom}
n = \log \left( \frac{1}{1-\eta} \right) \frac{\pi^{4} T^{4}\tau_s}{25920m^2}
\left(\nu_{\mathrm{max}}^{4}-\nu_{\mathrm{min}}^{4}\right)
\\ \times\
\left(a_{\mathrm{max}}^{3}-a_{\mathrm{min}}^{3}\right)
\left(\Omega_{\mathrm{max}}^{4}-\Omega_{\mathrm{min}}^{4}\right)
\left(\gamma_{\mathrm{max}}-\gamma_{\mathrm{min}}\right)
\,,
\end{multline}
where $\tau_s$ is the total timespan of the observation, $\mu$ is the maximal
mismatch, i.e. the maximal fractional loss in squared signal-to-noise ratio, and
$\eta$ is the covering probability, i.e. the probability of any particular point
in the space having a mismatch $<\mu$. The subscripts ``max'' and ``min''
identify the maximum and minimum values of the parameter ranges; see
Table~\ref{tab:par}.  The nominal sensitivity of our search to pulsations with a
fractional amplitude $A$ scales as
\begin{equation}
A = 2 \, M^{-1/4} \rho_{\Sigma}^{1/2} \langle\mathcal{N}\rangle^{-1/2} \,,
\end{equation}
where $\rho_{\Sigma}$ is the effective signal-to-noise ratio and
$\langle\mathcal{N}\rangle$ is the average number of photons in each segment.

For this work, the implementation of the above method used in~\cite{mes15}
underwent some optimisations. Instead of evaluating the derivatives
$\nu^{(m)}_{s}$ for every search frequency $\nu$, they are evaluated for a range
of $\nu$ values, i.e. $\nu^{(m)}_{s}(\nu) \approx \nu^{(m)}_{s}(\nu_0)$ where
$\nu \in [\nu_0 - \Delta\nu, \nu_0 + \Delta\nu]$.  This reduces the number of
computationally-expensive sine and cosine evaluations in Eq.~\eqref{eq:deriv}.
The range $\Delta\nu$ is chosen such that the difference
$|\nu^{(m)}_{0}(\nu) - \nu^{(m)}_{0}(\nu)|$ never exceeds half a grid spacing in
$\nu^{(m)}_{0}$.  The summation of power over segments was also vectorised using
single instruction, multiple data (SIMD) operations.  A factor of $\sim 7$
speed-up was gained by these optimisations.

\subsection{Search and Follow-Up Pipeline}\label{sec:search-follow-up}

For each LMXB, the setup of the search, defined by their variables
$(M, T, \mu)$ given above, is optimized so as to maximize the sensitivity
of the search at fixed computational cost, following the methodology of
\cite{pri12}.  The sensitivity of the search is
estimated using a variant of the analytic method derived in \cite{wet12};
throughout this paper we assume 1\% false alarm and 10\% false dismissal probabilities.
The value of $\eta$ was set to 90\%.  We chose to spend, per source, 24000
core-hours of the Atlas computer cluster of the Max Planck Institute for
Gravitational Physics, which at the time comprised chiefly of Intel
Xeon\footnote{E3-1220, 3.10GHz} cores.

The top 10 candidates from each search are then subjected to a follow-up search.
The parameter space for each follow-up search are centered on each candidate;
the range in $\nu$ was reduced to 1~Hz, and the ranges in $a$, $\Omega$, and
$\gamma$ are reduced to 10\% of their initial values.  The setups of each
follow-up search are optimised as per the initial search, with the requirement
that the minimum value of $T$ for each follow-up search must be twice that of
the initial search. The results of each follow-up search were then examined
manually; any search where noticeable peaks were seen in each parameter were
subject to additional follow-up searches following the above procedure. If the
signal is real, then this procedure should increase the signal-to-nose of the
candidate and highlight the presence of true pulsations.

\section{Validation}

\begin{table*}
\caption{Validation of Search Pipeline}
\centering
\begin{tabular}{lllcccc}
\hline
\hline
Source                            & Search Sensitivity & Parameter        & Recovered Value     & Actual Value        & Expected Error      & $|{\rm Recovered} - {\rm Actual}| / {\rm Error}$ \\
\hline
Software injection (strong)       & 1\%                & $A$              & 9.8\%               & 10\%                & 0.066\%             & 3.01                                             \\
                                  &                    & $\nu$ / Hz       & $\sn{5.988999}{2}$  & $\sn{5.989000}{2}$  & $\sn{1.273335}{-5}$ & 2.85                                             \\
                                  &                    & $a$ / lt-s       & $\sn{6.521037}{-2}$ & $\sn{6.500000}{-2}$ & $\sn{4.232632}{-5}$ & 4.97                                             \\
                                  &                    & $\Omega$ / rad/s & $\sn{7.103837}{-4}$ & $\sn{7.104380}{-4}$ & $\sn{1.674751}{-8}$ & 3.24                                             \\
                                  &                    & $\gamma$ / rad   & 2.797675            & 2.796213            & 0.000649            & 2.25                                             \\
\hline
Software injection (moderate)     & 1\%                & $A$              & 3.2\%               & 3.3\%               & 0.069\%             & 1.93                                             \\
                                  &                    & $\nu$ / Hz       & $\sn{5.989001}{2}$  & $\sn{5.989000}{2}$  & $\sn{3.940856}{-5}$ & 3.14                                             \\
                                  &                    & $a$ / lt-s       & $\sn{6.513522}{-2}$ & $\sn{6.500000}{-2}$ & $\sn{1.310070}{-4}$ & 1.03                                             \\
                                  &                    & $\Omega$ / rad/s & $\sn{7.103238}{-4}$ & $\sn{7.104380}{-4}$ & $\sn{5.189617}{-8}$ & 2.20                                             \\
                                  &                    & $\gamma$ / rad   & 2.792518            & 2.796213            & 0.002011            & 1.83                                             \\
\hline
Software injection (borderline)   & 1\%                & $A$              & 1.1\%               & 1.1\%               & 0.089\%             & 0.24                                             \\
                                  &                    & $\nu$ / Hz       & $\sn{5.990859}{2}$  & $\sn{5.989000}{2}$  & $\sn{1.157234}{-4}$ & 1600                                             \\
                                  &                    & $a$ / lt-s       & $\sn{5.647940}{-2}$ & $\sn{6.500000}{-2}$ & $\sn{3.774911}{-4}$ & 22.5                                             \\
                                  &                    & $\Omega$ / rad/s & $\sn{7.236696}{-4}$ & $\sn{7.104380}{-4}$ & $\sn{1.724539}{-7}$ & 76.7                                             \\
                                  &                    & $\gamma$ / rad   & 3.205199            & 2.796213            & 0.006683            & 61.1                                             \\
\hline
Software injection (subthreshold) & 1\%                & $A$              & 1\%                 & 0.37\%              & 0.26\%              & 2.48                                             \\
                                  &                    & $\nu$ / Hz       & $\sn{6.033609}{2}$  & $\sn{5.989000}{2}$  & $\sn{1.205446}{-4}$ & 37000                                            \\
                                  &                    & $a$ / lt-s       & $\sn{6.797894}{-2}$ & $\sn{6.500000}{-2}$ & $\sn{3.973361}{-4}$ & 7.49                                             \\
                                  &                    & $\Omega$ / rad/s & $\sn{7.110951}{-4}$ & $\sn{7.104380}{-4}$ & $\sn{1.508134}{-7}$ & 4.35                                             \\
                                  &                    & $\gamma$ / rad   & 5.850171            & 2.796213            & 0.005844            & 522                                              \\
\hline
Blind injection challenge         & 0.74\%             & $A$              & 1.4\%               & 1.2\%               & 0.05\%              & 3.03                                             \\
                                  &                    & $\nu$ / Hz       & $\sn{2.873471}{2}$  & $\sn{2.873470}{2}$  & $\sn{4.785520}{-5}$ & 2.71                                             \\
                                  &                    & $a$ / lt-s       & $\sn{8.113484}{-1}$ & $\sn{8.110000}{-1}$ & $\sn{9.445180}{-4}$ & 0.36                                             \\
                                  &                    & $\Omega$ / rad/s & $\sn{2.493601}{-4}$ & $\sn{2.493327}{-4}$ & $\sn{6.910871}{-9}$ & 3.96                                             \\
                                  &                    & $\gamma$ / rad   & 3.631982            & 3.676000            & 0.001164            & 37.8                                             \\
\hline
SAX J1808.4--3658                 & 0.53\%             & $A$              & 7.3\%               & 7.8\%               & 0.025\%             & 20.3                                             \\
                                  &                    & $\nu$ / Hz       & $\sn{4.009751}{2}$  & $\sn{4.009752}{2}$  & $\sn{7.716828}{-6}$ & 5.41                                             \\
                                  &                    & $a$ / lt-s       & $\sn{6.296443}{-2}$ & $\sn{6.280800}{-2}$ & $\sn{3.139776}{-5}$ & 4.98                                             \\
                                  &                    & $\Omega$ / rad/s & $\sn{8.668372}{-4}$ & $\sn{8.667472}{-4}$ & $\sn{6.426050}{-9}$ & 14.0                                             \\
                                  &                    & $\gamma$ / rad   & 2.479961            & 2.539119            & 0.000498            & 118                                              \\
\hline
IGR J00291+5934, section 1        & 0.99\%             & $A$              & 8\%                 & 10\%                & 0.066\%             & 30.7                                             \\
                                  &                    & $\nu$ / Hz       & $\sn{5.988922}{2}$  & $\sn{5.988921}{2}$  & $\sn{1.565229}{-5}$ & 8.38                                             \\
                                  &                    & $a$ / lt-s       & $\sn{6.358381}{-2}$ & $\sn{6.498700}{-2}$ & $\sn{5.199202}{-5}$ & 26.9                                             \\
                                  &                    & $\Omega$ / rad/s & $\sn{7.108983}{-4}$ & $\sn{7.104406}{-4}$ & $\sn{2.109826}{-8}$ & 21.6                                             \\
                                  &                    & $\gamma$ / rad   & 1.420043            & 1.477473            & 0.000817            & 70.2                                             \\
\hline
IGR J00291+5934, section 2        & 1.3\%              & $A$              & 5.9\%               & 6\%                 & 0.12\%              & 1.17                                             \\
                                  &                    & $\nu$ / Hz       & $\sn{5.988918}{2}$  & $\sn{5.988921}{2}$  & $\sn{4.185645}{-5}$ & 5.72                                             \\
                                  &                    & $a$ / lt-s       & $\sn{6.393486}{-2}$ & $\sn{6.498700}{-2}$ & $\sn{1.391573}{-4}$ & 7.56                                             \\
                                  &                    & $\Omega$ / rad/s & $\sn{7.102692}{-4}$ & $\sn{7.104406}{-4}$ & $\sn{8.578632}{-8}$ & 1.99                                             \\
                                  &                    & $\gamma$ / rad   & 0.380340            & 0.423260            & 0.002176            & 19.7                                             \\
\hline
IGR J00291+5934, section 3        & 1.5\%              & $A$              & 1.6\%               & 1\%                 & 0.18\%              & 3.31                                             \\
                                  &                    & $\nu$ / Hz       & $\sn{5.988974}{2}$  & $\sn{5.988921}{2}$  & $\sn{1.088178}{-4}$ & 49.3                                             \\
                                  &                    & $a$ / lt-s       & $\sn{7.467071}{-2}$ & $\sn{6.498700}{-2}$ & $\sn{3.669491}{-4}$ & 26.3                                             \\
                                  &                    & $\Omega$ / rad/s & $\sn{7.002556}{-4}$ & $\sn{7.104406}{-4}$ & $\sn{4.621755}{-8}$ & 220                                              \\
                                  &                    & $\gamma$ / rad   & 3.357818            & 5.418901            & 0.004914            & 419                                              \\
\hline
\end{tabular}
\label{tab:valid}
\end{table*}

Prior to analyzing the data of the eleven LMXBs, we performed several tests to
validate our data preparation, search pipeline, and sensitivity estimates.
These expand upon similar tests of the search pipeline in \cite{mes15}.

To check the ability of our search pipeline to recover signals of varying strength, we prepared simulated datasets spanning $\sim 134$~ks, with $\sim 25.7$~ks of on-source time.
These datasets contained a randomly-generated background of $\sim\sn{2.18}{6}$ photons and a simulated signal, following Eq.~\eqref{eq:phase}, with fractional amplitudes of 10\%, 3.3\%, 1.1\%, and 0.37\%.
The datasets were searched using a search with an estimated sensitivity of 1\%; relative to this sensitivity the four injections correspond to strong, moderate, borderline, and subthreshold strengths respectively.
Table~\ref{tab:valid} compares the parameters of the loudest candidate recovered from each search against the actual parameters of the injected signal.
We see that, aside from the subthreshold case, our recovered parameters are mostly in good agreement with their actual values.
The difference between recovered and actual fractional amplitudes $A$ are within a few factors of the expected error given by the standard deviation of the detection statistic \citep[Eqs.~10]{mes15}.
The differences between recovered and actual parameters $\nu$, $a$, $\Omega$, and $\gamma$ are, with a few expections, within a few factors of the expected error given by the Cram\'er--Rao lower bound.

To further test our data preparation and search pipeline, we performed the following blind injection challenge.
One author prepared a simulated outburst with the same length and number of photons
as found in 4U 1323--619, and injected a fake signal in the simulated data.
The data was then blindly searched by another author who was unaware of the true parameters of the fake signal.
The search covered a wide parameter space of $\nu \in 100$--1000~Hz, $a \in 0.810$--0.817~lt-s, $\Omega \in (2.4896$--$\sn{2.4957)}{-4}$~rad/s, and $\gamma \in 0$--$2\pi$, and had a sensitivity of 0.74\%.
As seen in Table~\ref{tab:valid}, the signal was recovered at a fractional amplitude and parameters broadly consistent with the expected errors.

To confirm that our search pipeline is able to find pulsations from real pulsars, we then searched data from the known AMXPs SAX J1808.4--3658 (using the 1998 outburst) and IGR J00291+5934 (the 2008 outburst), using data recorded by \textit{RXTE} and prepared using the same processing described in Section~\ref{sec:obs}. 
Data from IGR J00291+5934 was split into 3 sections within which pulsations are recorded at fractional amplitudes of 10\%, 6\%, and 1\%\footnote{In section 3, pulsations are observed at 2\% in 3 out of 13 contiguous data stretches; no pulsations are observed in the remaining 10 stretches. The 3 stretches comprise $\sim 50$\% of the photons accumulated during section 3, so we take the fractional amplitude over the entire section to be 1\%.} respectively.
For strong pulsations (SAX J1808.4--3658 and IGR J00291+5934, section 1) our recovered fractional amplitudes are slightly less than expected; for SAX J1808.4--3658 $A = 7.3$\% recovered against 7.8\% expected, and for IGR J00291+5934, section 1, $A = 8$\% recovered against 10\% expected.
Nevertheless we clearly recover the known pulsar signal, and at the correct parameters.
The same statement is true of the weaker 6\% pulsations in IGR J00291+5934, section 2; for IGR J00291+5934, section 3, the 1\% pulsations are below the 1.5\% sensitivity of the search, and therefore we do not expect a detection.

Finally, to double-check our sensitivity estimation and optimisation procedure, we reproduce the search for the 3rd outburst of Aql X-1 performed in \cite{mes15}.
The search performed in this paper covered the same search parameter space as \cite{mes15} using a setup with $M = 250$, $T = 275$~s, and $\mu = 0.0126$; for comparison \cite{mes15} used $M = 258$, $T = 256$~s, and $\mu = 0.1$.
The sensitivity of the search was estimated, using the procedure described in Section~\ref{sec:search-procedure}, to be 0.24\%.
This is consistent with the 0.26\% estimated by \cite{mes15} for the sensitivity of their analysis, and is expected given that the search setups are very similar (apart from the smaller $\mu$ used in this paper).


\section{Results}\label{sec:res}

\begin{table*}
\caption{Search Parameters and Upper Limits on Pulsed Fraction}
\centering
\begin{tabular}{lccccccccc}
\hline
\hline
Source                & \multicolumn{5}{c}{Search}                                  & \multicolumn{4}{c}{Follow-Up}                                                                      \\
                      & $T$ (s) & $M$  & $\mu$  & $n$            & $A^{\rm UL90\%}$ & $T_{\rm FU}$  & $\mu_{\rm FU}$  & $n_{\rm FU}$                   & $A^{\rm UL90\%}_{\rm FU~Best}$  \\
\hline
4U 1323--619 (wide)   & 71      & 4600 & 0.724  & $\sn{1.1}{12}$ & 1.6\%            & 530           & 0.0768--0.498   & $\sn{4.2}{10}$--$\sn{1.3}{11}$ & 0.65\%                          \\
4U 1323--619 (narrow) & 210     & 1552 & 0.716  & $\sn{2.6}{12}$ & 1.0\%            & 530           & 0.0123--0.123   & $\sn{7.6}{9}$--$\sn{1.4}{10}$  & 0.64\%                          \\
4U 1456--32 (Cen X-4) & 183     & 315  & 0.0166 & $\sn{6.0}{12}$ & 6.0\%            & 1236--1522    & 0.0109--0.0224  & $\sn{1.4}{9}$--$\sn{7.2}{10}$  & 2.9\%                           \\
4U 1543--624          & 54      & 543  & 0.0928 & $\sn{8.7}{12}$ & 0.84\%           & 109           & 0.0675--0.0864  & $\sn{4.3}{8}$--$\sn{1.3}{9}$   & 0.63\%                          \\
4U 1608--52           & 256     & 152  & 0.126  & $\sn{1.0}{13}$ & 0.18\%           & 512           & 0.0118--0.0366  & $\sn{1.8}{11}$--$\sn{7.1}{11}$ & 0.14\%                          \\
4U 1636--53           & 347     & 142  & 0.0132 & $\sn{1.3}{9}$  & 0.22\%           & 695           & 0.0129--0.0186  & $\sn{5.5}{5}$--$\sn{1.3}{7}$   & 0.17\%                          \\
XTE J1710--28         & 120     & 4694 & 0.722  & $\sn{1.1}{12}$ & 1.6\%            & 590           & 0.0438--0.257   & $\sn{2.6}{10}$--$\sn{1.1}{11}$ & 0.78\%                          \\
4U 1735--44           & 331     & 193  & 0.68   & $\sn{8.1}{12}$ & 0.21\%           & 663           & 0.05--0.467     & $\sn{8.7}{8}$--$\sn{7.5}{9}$   & 0.14\%                          \\
XTE J1739--2859       & 20      & 4797 & 0.724  & $\sn{1.0}{12}$ & 1.6\%            & 40--556       & 0.0535--0.713   & $\sn{1.4}{11}$--$\sn{7.3}{11}$ & 0.48\%                          \\
4U 1746--37           & 119     & 3501 & 0.723  & $\sn{1.4}{12}$ & 0.62\%           & 238--796      & 0.0137--0.158   & $\sn{4.8}{10}$--$\sn{1.5}{11}$ & 0.28\%                          \\
XTE J2123--058        & 198     & 321  & 0.693  & $\sn{9.7}{12}$ & 0.57\%           & 396--594      & 0.0168--0.0274  & $\sn{2.3}{11}$--$\sn{4.5}{11}$ & 0.34\%                          \\
4U 2129+12 (AC 211)   & 570     & 578  & 0.703  & $\sn{6.0}{12}$ & 0.58\%           & 1800          & 0.0322--0.0639  & $\sn{2.4}{11}$--$\sn{3.0}{11}$ & 0.35\%                          \\
\hline
\end{tabular}
\label{tab:ul}
\end{table*}

We find no new pulsations in the eleven LMXBs considered in this work.  In
Table~\ref{tab:ul} we report the 90\% confidence level upper limits for the full
parameter space explored ($A^{\rm UL90\%}$) along with the best upper limits
from the follow-up search on candidates ($A^{\rm UL90\%}_{\rm FU~Best}$).  The
best upper limits are of $\approx 0.2$\% for the sources 4U 1608--52, 4U
1735--44, and 4U 1636--53. For 4U 1608--52 we find a maginally significant
candidate during our full parameter space search, with a fractional amplitude of
$A = 0.17$\% and with parameters $\nu = 617.18$~Hz, $P = 45253$~s,
$a = 0.72737$~lt-s and $\gamma = 0.92823$~rad. The candidate was not found,
however, when folding the data coherently (using the code
PRESTO;~\citealt{ran02,ran11}) and exploring a small parameter space around the
best candidate. A search of a different \textit{RXTE} dataset from 4U 1608--52
also revealed no pulsations at the same parameters.

The current upper limits are close to the best possible value that can be
achieved with current datasets and computational resources.  These results are
similar in order of magnitude to what previously found in Aql
X-1~\citep{mes15}.  Some upper limits represent an improvement of a factor of 10
with respect to previous upper limits either published in the literature or
obtainable by simply looking at short-length\footnote{Assuming that a signal is
  present in a certain binary with a given orbit, the maximum time to keep all
  the power in one Fourier frequency bin when doing simple power spectra,
  without orbital corrections, is $2\pi a / P \nu$ \citep{van88}.} power
spectra.


\section{Discussion}\label{sec:dis}

Together with other previous pulse searches in LMXBs~\citep{vau94, dib04,
  mes15}, the lack of weak pulsations in these eleven LMXBs implies that weak
pulses are not present in most LMXBs. What was found in Aql X-1 (with upper
limits of $0.27$\% on the pulsed fraction) therefore cannot be considered an
anomalous behavior but rather the norm. The small values of the upper limits on
the pulsed fractions imply that, if we are able to see the surface of the
neutron stars, the emission pattern originating must be extraordinarily uniform
with no obvious asymmetries.

The various mechanisms that might be responsible for this behaviour have been
extensively discussed in the literature and we refer to \citet{mes15} for
details. Here we note that different pieces of evidence coming from a number of
independent studies seem to be converging towards the lack of a magnetosphere
around most accreting neutron stars in LMXBs.  Beside the negative results of
deep pulse searches (27 LMXBs with published results so far, including this
work), the most important ones are:
\begin{itemize}
\item the aperiodic variability of AMXPs shows shifted correlations of power
  spectral components with respect to non-pulsating atoll sources~\citep{van05};
\item the quiescent LMXB Cen X-4 has shown no evidence for pulsations and
  modeling of its spectral behavior favour the presence of a radiatively
  inefficient accretion flow rather than a propeller (which would be expected
  if a magnetosphere were present; \citealt{dan15});
\item the existence of two sub-populations in LMXBs \citep{pat17}, with the easiest
  possibility being that no magnetosphere is present in some LMXBs;
\item very different behavior of burst oscillations in AMXPs \& non-pulsating
  LMXBs, with the former showing pulse phase locking between accretion and
  nuclear powered pulsations~\citep{wat08,cav11}, burst oscillation frequency
  overshooting the spin frequency~\citep{cha03}, burst oscillations present in
  all bursts (and only sometimes in non-pulsating LMXBs) and a strong harmonic
  content vs. little to no harmonic content in non-pulsating
  sources~\citep{str03,wat09b};
\item the lack of short intermittent pulse episodes in 40 LMXBs (Algera \& Patruno
  in prep.);
\item exponentially decreasing accretion torques in the intermittent AMXP HETE
  J1900.1--2455 compatible with a decreasing magnetosphere
  strength~\citep{pat12c};
\item the aperiodic variability of the intermittent source HETE J1900.1--2455
  behaves as non-pulsating atoll sources rather than AMXPs~\citep{pat17b}.
\end{itemize}

It seems therefore plausible to suggest that the lack of pulsations in LMXBs can
be ascribed to a weak/no magnetosphere. This scenario comes of course with
shortcomings since a number of other observational results would still not be
easily explained. For example, a weak magnetosphere would not justify why Aql
X-1 has shown such short ($\approx 150$ s) but strong ($\approx 6.5$\% pulsed
fraction) pulse episode. It is also difficult to understand what causes the
weakness of the magnetosphere. Initial suggestions~\citep{bis74, cum01} focused
on the mass accretion rate, which was proposed to be higher in non pulsating
systems. However, the observational evidence now suggests that the mass
accretion rate cannot be the only cause for the lack of pulsations since these
are not seen also in some faint systems too~\citep{pat10p}. Furthermore, there is a strong tension between the lack of pulsations in Aql X-1 and the recent claim that a relatively strong magnetosphere is present around this system, with a disk truncated at a few tens of kilometer from the neutron star surface~\citep{lud17}. Indeed any magnetosphere around this system requires a strong fine tuning of the parameters to explain the lack of pulses (see \citealt{mes15} for a discussion).

Another result whose explanation remains problematic is that the
intermittent AMXP SAX J1748.9--2021 was observed first as a non-pulsating atoll
source (in 1998), then it turned into an intermittent AMXP (in 2001, 2005 \&
2009; \citealt{alt08,Patruno09, pat10e}) and then it became a persistent AMXP in
2015~\citep{san16}. In this case therefore the neutron star magnetosphere, if absent in 1998, must have re-emerged on a relatively short timescale for a reason not completely clear. 

There is of course also the possibility that some of our underlying assumptions
used in the pulse search were not correct. For example our upper limits are
valid only if the weak pulsations are always present. If they are appearing
intermittently then the upper limits we calculate might be off by a large amount
(that depends on the fraction of time the pulsations are on).  A second
assumption is that the true orbital parameters really lie within the range
explored in this search. In particular the orbital periods determined from
optical observations are often affected by systematics and it might be possible
that some unaccounted effects occur also in the determination of those here
selected (see e.g., \citealt{pat17c} for an overview of such effects). However, it is difficult to believe that none of the eleven (sometimes
very conservative) ranges chosen contain at least one of the true orbital
periods.

Finally, drifting pulse phases (often in response to X-ray flux variations) have
been observed in basically all AMXPs~\citep{har08,pat09f,pat12} and this effect
has been interpreted as a moving hot spot on the neutron star surface. The drift
occurs on timescales of hours/days but it can be as short as few
minutes~\citep{pat12c}. However, if shorter and varying timescales are involved
for the hot spot motion in most LMXBs (with the AMXPs being the sources where
the motion is the slowest) it is possible to lose the coherence of the signal
even if a relatively strong magnetosphere is present. This possibility remains
speculative at the moment, but a better understanding of the physical mechanism
inducing the hot spot motion might help to clarify the issue.

\acknowledgments AP acknowledges support from a Netherlands Organization for
Scientific Research (NWO) Vidi Fellowship.
The pulse searches were performed on the Atlas computer cluster of the
Max Planck Institute for Gravitational Physics.

\bibliographystyle{apj}
\bibliography{biblio}

\end{document}